\documentclass[12pt,twoside,english]{article}
\usepackage[T1]{fontenc}
\usepackage[latin1]{inputenc}
\usepackage{geometry}
\usepackage{babel}
\usepackage{graphics}

\begin{document}
\def\AEF{A.E. Faraggi}
\def\NPB#1#2#3{{\it Nucl.\ Phys.}\/ {\bf B#1} (#2) #3}
\def\PLB#1#2#3{{\it Phys.\ Lett.}\/ {\bf B#1} (#2) #3}
\def\PRD#1#2#3{{\it Phys.\ Rev.}\/ {\bf D#1} (#2) #3}
\def\PRL#1#2#3{{\it Phys.\ Rev.\ Lett.}\/ {\bf #1} (#2) #3}
\def\PRT#1#2#3{{\it Phys.\ Rep.}\/ {\bf#1} (#2) #3}
\def\MODA#1#2#3{{\it Mod.\ Phys.\ Lett.}\/ {\bf A#1} (#2) #3}
\def\IJMP#1#2#3{{\it Int.\ J.\ Mod.\ Phys.}\/ {\bf A#1} (#2) #3}
\def\nuvc#1#2#3{{\it Nuovo Cimento}\/ {\bf #1A} (#2) #3}
\def\RPP#1#2#3{{\it Rept.\ Prog.\ Phys.}\/ {\bf #1} (#2) #3}
\def\APJ#1#2#3{{\it Astrophys.\ J.}\/ {\bf #1} (#2) #3}
\def\APP#1#2#3{{\it Astropart.\ Phys.}\/ {\bf #1} (#2) #3}
\def\etal{{\it et al\/}}

\begin{titlepage}

\setcounter{page}{1}
\rightline{}
\vfill
\begin{center}
 {\Large \bf Identifying String Relics at AUGER?
\footnote{Presented by Claudio Corian\`o at String Phenomenology 2003, Durham, U.K. July 2003}}

\vfill
\vfill
{\large Alessandro Cafarella and Claudio Corian\`{o}}

\vspace{.12in}
 {\it  Dipartimento di Fisica, Universit\`{a} di Lecce \\
 and INFN Sezione di Lecce \\ Via Arnesano 73100 Lecce, Italy}

\vspace{.075in}

\end{center}
\vfill

\begin{abstract}
The identification of string relics, 
or of other very massive states, at forthcoming ultra high energy 
cosmic rays experiments, requires a good reconstruction of the 
main properties of the extensive air showers produced by the 
collision of the primary protons with the atmosphere. 
In particular, the current hadronization  
models used in the simulations need to incorporate 
possible new interactions. 
We briefly discuss these aspects 
and then proceed by describing some of the observables which 
characterize the atmospheric shower. The linear growth of the multiplicities 
- as a function of the energy - for all the main particles in the shower 
can play an interesting role in an attempt to identify channels of 
missing energy due to a plausible dark matter component.

\end{abstract}
\smallskip

\end{titlepage}

\setcounter{footnote}{0}

% ========================= DEFINITIONS ===================================
\def\beq{\begin{equation}}
\def\eeq{\end{equation}}
\def\beqn{\begin{eqnarray}}
\def\eeqn{\end{eqnarray}}
\def\ba{\begin{eqnarray}}
\def\ea{\end{eqnarray}}
\def\ie{{\it i.e.}}
\def\eg{{\it e.g.}}
\def\half{{\textstyle{1\over 2}}}
\def\nicefrac#1#2{\hbox{${#1\over #2}$}}
\def\third{{\textstyle {1\over3}}}
\def\quarter{{\textstyle {1\over4}}}
\def\m{{\tt -}}

\def\p{{\tt +}}

\def\slash#1{#1\hskip-6pt/\hskip6pt}
\def\slk{\slash{k}}
\def\GeV{\,{\rm GeV}}
\def\TeV{\,{\rm TeV}}
\def\y{\,{\rm y}}
\def\ds{\slash}
\def\l{\langle}
\def\r{\rangle}
\def\xprime{x^{\prime}}
\def\xprimetwo{x^{\prime\prime}}
\def\zprime{z^{\prime}}
\def\xprimbar{\overline{x}^\prime}
\def\xprim2bar{\overline{x}^{\prime\prime}}
\def\ptbold{\mbox{\boldmath$p$}_T}
\def\ktbold{\mbox{\boldmath$k$}_T}
\def\ktboldbar{\mbox{\boldmath$\overline{k}$}_T}
\def\beq{\begin{equation}}
\def\eeq{\end{equation}}
\def\tr{{\bf tr}}
\def\P{P^\mu}
\def\Pb{\overline{P}^\mu}
\def\BOX#1#2#3#4#5{\hskip#1mm\raisebox{#2mm}[#3mm][#4mm]{$#5$}}
\def\VBOX#1#2{\vbox{\hbox{#1}\hbox{#2}}}

\setcounter{footnote}{0}
\newcommand{\beqa}{\begin{eqnarray}}
\newcommand{\eeqa}{\end{eqnarray}}
\newcommand{\eps}{\epsilon}

\pagestyle{plain}
\setcounter{page}{1}

\section{Introduction}
The search for new physics in high energy 
cosmic rays has intensified over the 
last few years \cite{uhecr,auger,euso} (see \cite{la} for a review). 
In fact, several experimental collaborations 
have provided interesting new data in this direction, 
even though, at the moment, there is not 
enough evidence to conclusively assess the presence (or the exclusion)
of a Greisen-Zatsepin-Kuzmin (or GZK) cutoff \cite{gzk} ($\sim 10^{19}$ eV) on the basis 
of these sets of data alone. 

It is widely expected that the presence of this cutoff in the 
- almost structureless - fast falling (in energy) 
inclusive cosmic ray spectrum is due to the interaction of the 
primary particles with the cosmic background radiation.   
The presence of a cutoff implies that the sources should 
be inside a sphere of approximately $50$ Mpc (GZK distance) in radius. Events above 
the cutoff (also termed Ultra High Energy Cosmic Rays or UHECR), 
which have been repeatedly reported by several experimental 
collaborations in the last fourty years or so,
if they are experimentally confirmed, should therefore point directly to their 
sources located within this radial GZK distance from our galaxy. 
Since within this distance there is no homogeneity in the distribution 
of sources, if we assume a ``traditional'' or ``bottom-up'' mechanism 
of acceleration, such as Fermi acceleration and 
variants thereof, 
we should detect clustering in the events. If clustering is not observed then 
we need to look for alternative (non traditional) explanations, 
with very interesting consequences.

It has been thought for some time 
that  new physics is involved 
in the explanation of the UHECR. 
From a theoretical perspective, the confirmation of 
the existence of a systematic violation of the cutoff 
can be considered as a possible signal of new physics. 

There are many interesting suggestions as for the nature of these events. One possibility is the well known Z-burst mechanism (see \cite{weiler} and refs. 
therein), based on the idea of a ultra high energy neutrino (UHE) hitting a relic 
anti-neutrino. The corresponding s-channel amplitude may resonate on the Z gauge boson with a larger cross section. The typical energy of the 
primary protons generated by this mechanism 
is around the GZK cutoff and is fine-tuned by a neutrino 
mass which is in agreement with current estimated values. 
This suggestion evades the cutoff due to the weakly interacting nature of 
the primary (a neutrino). We recall that the possibility of a direct interaction of neutrinos scattering off nucleons in the atmosphere is ruled out by the fact that 
such interaction should take place uniformly in the 
atmosphere, and it doesn't. 

Other, more radical suggestions 
call for a modification of fundamental physics \cite{explanations}. 
We should mention that 
several re-analysis of the data by the HiRes \cite{hires} 
and the AGASA \cite{agasa} collaborations 
also add up to the dispute about the very existence of the cutoff, while 
other proposals call for dark matter candidates (see \cite{sarkar} for 
a summary). 
There are various suggestions in this last direction and there is enough room, 
allowed both from cosmological constraints and particle physics constraints, 
to have viable particle candidates as origin of the UHECR flux 
without running into some narrow corners of parameter space.    

It is expected that the AUGER observatory \cite{auger} in 
Argentina will be able to provide a definitive answer to this issues, 
while the construction of a separate (AUGER NORTH) site is also 
under consideration. 

\section{String Relics}

\begin{figure}[t]
{\centering {\resizebox*{14cm}{!}{\rotatebox{-90}{\includegraphics{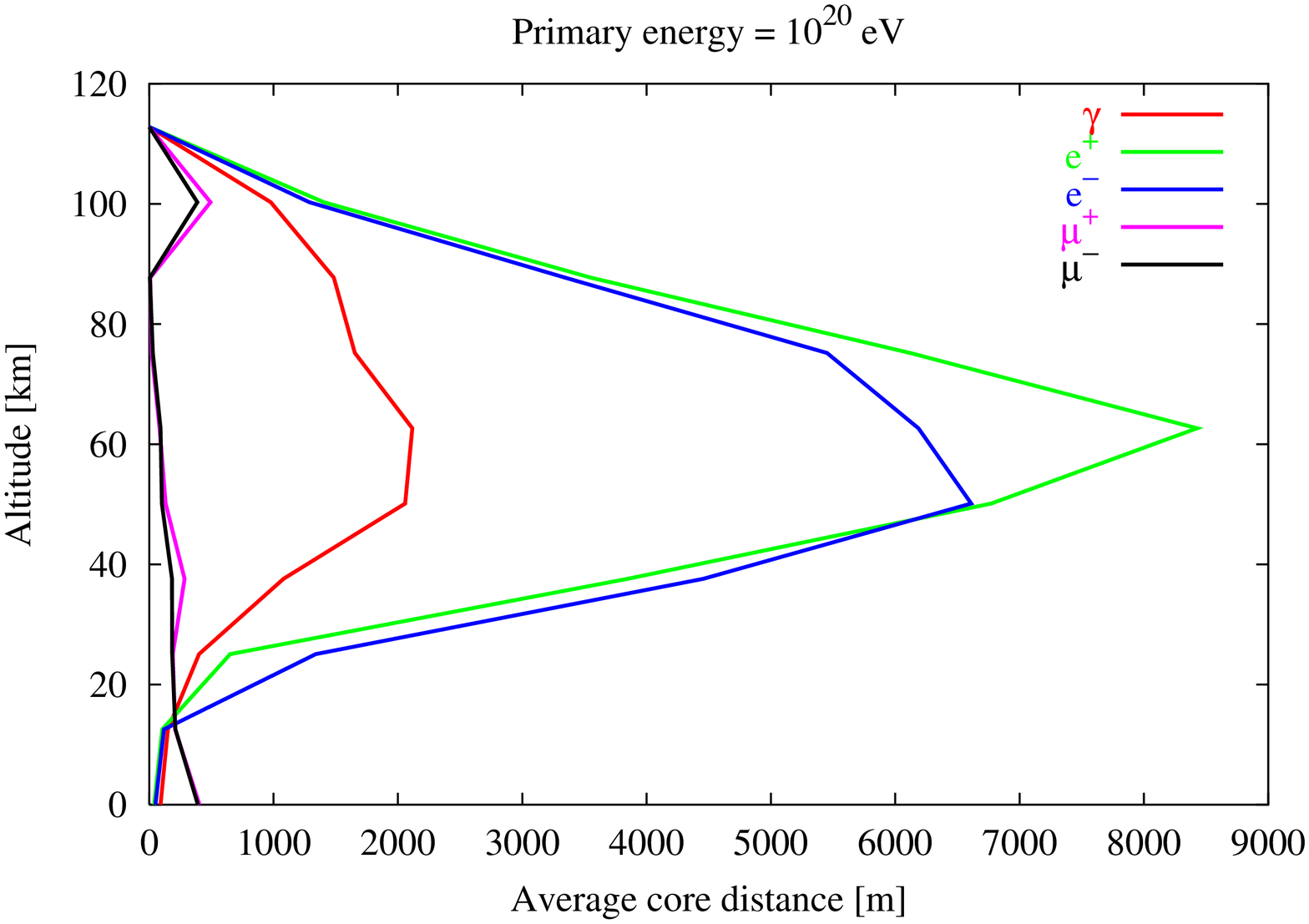}}}} \par}
\caption{Average core distancies of photons, \protect\( e^{\pm }\protect \),
\protect\( \mu ^{\pm }\protect \) at various levels of observations
for a primary energy of \protect\( 10^{19}\protect \) eV. The first
impact is forced to occur at the top of the atmosphere.}
\label{ninth}
\end{figure}

As we have mentioned, one of the possible solution to the problem of the 
origin of the UHECR events is obtained by assuming that the primary proton 
spectrum
is generated by the decay of long-lived super-heavy states whose mass is in the $10^{12-15}$ GeV range (see \cite{ccf,cfp} and refs. therein). 

Superstring theory can naturally account for the meta-stable states by a 
stabilization mechanism due to the breaking of the non-abelian gauge symmetries by Wilson lines. The mechanism \cite{ccf} gives rise 
to states in the string spectrum which carry standard charge under the 
Standard Model, but fractional charge under an additional
$U(1)_{Z^\prime}$ gauge group, which can be regarded as a rather 
generic consequence of string unification. 

In this case the GZK cutoff is evaded assuming that these 
relics are distributed within GZK distance from us.
The mass required for the meta-stable state, whose lifetime is between 
$10^{17}-10^{27}$ sec, is about $10^{12-13}$ GeV, while their abundancy 
($\sim 5\times 10^{-11}$) is constrained by the observed flux of the UHECR events. We remark that the spectrum of the primary hadrons/leptons generated 
by the decay can be calculated - in some approximation - 
using standard renormalization group tools 
(in the vacuum) \cite{fragfun}, while their energy can be comparable 
with the GZK energy. On the basis of these results one 
can reasonably assess that long lived meta-stable particles 
can be candidates for the origin of UHECR events. 

\begin{figure}[t]
{\centering {\resizebox*{14cm}{!}{\rotatebox{-90}{\includegraphics{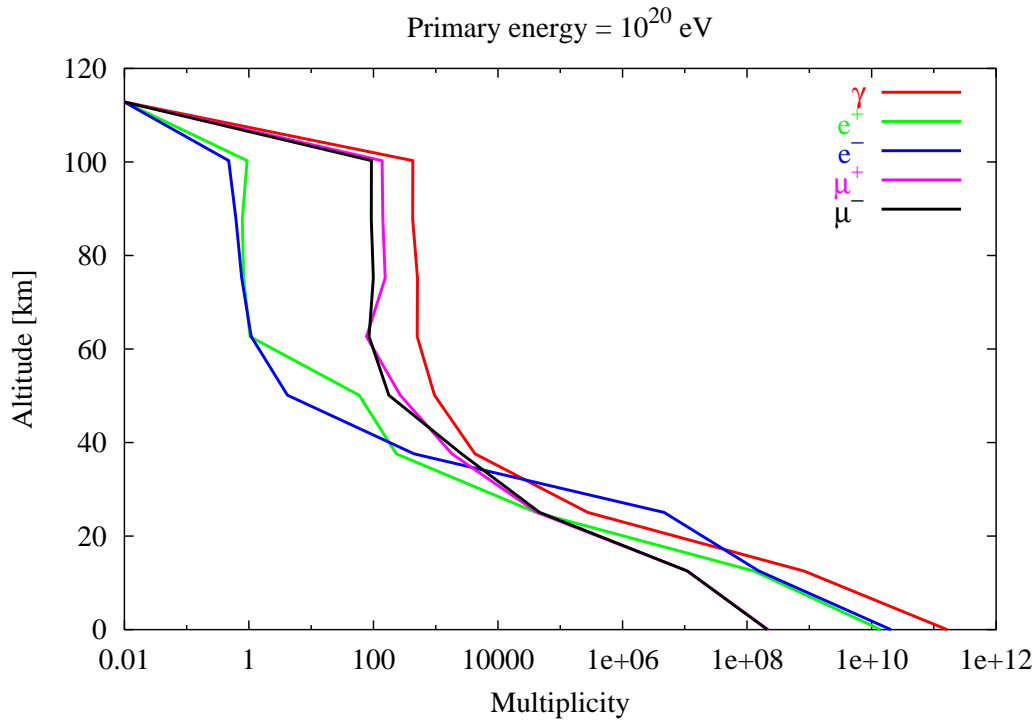}}}} \par}
\caption{Multiplicities of photons, \protect\( e^{\pm }\protect \), \protect\( \mu ^{\pm }\protect \)
at various levels of observations for a primary energy of \protect\( 10^{20}\protect \)
eV. The first impact is forced to occur at the top of the atmosphere.}
\label{sixth}
\end{figure}

\section{Descriptions/Modeling}

The analysis of the structural properties of the shower after the 
impact of the primary cosmic ray with the atmosphere is the 
main objective in the work of the 
experimental collaborations and relies heavily on the use of 
complex Monte Carlo programs, such as CORSIKA \cite{corsika}. 
Usually, these Monte Carlo have, on the other hand, 
to rely on the performance of special event generators 
to simulate the first impact, which use data on inclusive cross sections 
and partial cross sections 
at lower energies extrapolated up to very high energies, around the 
cutoff. Both a diffractive (dominant) and a non-diffractive component 
are usually included in these models, to which a minijet cross 
section is also added up. These routines 
are called up by the main program repeatedly, both for scattering 
and fragmentation. 
At a second stage, iterativly, the main code simulates the development 
of the air shower along the atmosphere, going exhaustively through 
all the possible channels and tracking down the particles which have been generated 
at each interaction. The cascade evolves as a {\em fat tree} 
and is computationally very expensive. At the GZK cutoff 
one soon runs into memory problems and a direct simulation 
needs therefore special algorithms \cite{thinning} to render the analysis 
possible. 

Having said this, we come to illustrate other work in the formation of 
the fragmenting spectrum {\it in the vacuum} which also gives some 
indication on the multiplicity structure 
of the various channels open to the fragmenting primary.

These attempts \cite{fragfun} have been based on the use of standard renormalization group 
(RG) equations, incorporating at times the resummation of small-x 
effects, for instance using color coherence in the Modified Leading 
Logarithmic approximation, in the study of the final spectrum 
of the heavy particle decaying. 
We remark that these second attempts usually neglect the role played by the passage of the 
primaries in the atmosphere in the determination of the final signal on the 
plane of the detector. In fact, the rearrangement of the spectrum of the final hadrons induced by the mixing of the anomalous dimensions in the RG 
evolution, as first suggested in \cite{cc}, 
due -for instance- to supersymmetry, gives some indication on 
the presence of a new component, but should be folded with the ordinary atmospheric simulations
in order to produce more reliable results \cite{cacf}. 
We simply do not have, at this time, any analysis which is, from this 
perspective, complete enough to be useful for the experiments
and more work is needed in this direction.
 
In regard to the study of horizontal air showers, much less 
is known at a quantitative level, compared to air showers induced by primary hadrons, since the interaction of neutrinos with the atmosphere is not included in simulation programs such as CORSIKA.
We should mention that 
the column of air above 
the detector can act as a large calorimeter, 
thereby increasing the rate for neutrino interaction considerably.
It is important though to remark that horizontal air showers are  
depleted of the electromagnetic component\footnote{We thank J. Knapp 
for comments on this point} and have mostly a muon component which 
makes the shape of the cascade {\it distinct} from the hadron initiated shower.
Improvements along these lines will be very important in order 
to reconstruct these near-horizontal air showers.   

\begin{figure}[t]
{\centering \resizebox*{14cm}{!}{\rotatebox{-90}{\includegraphics{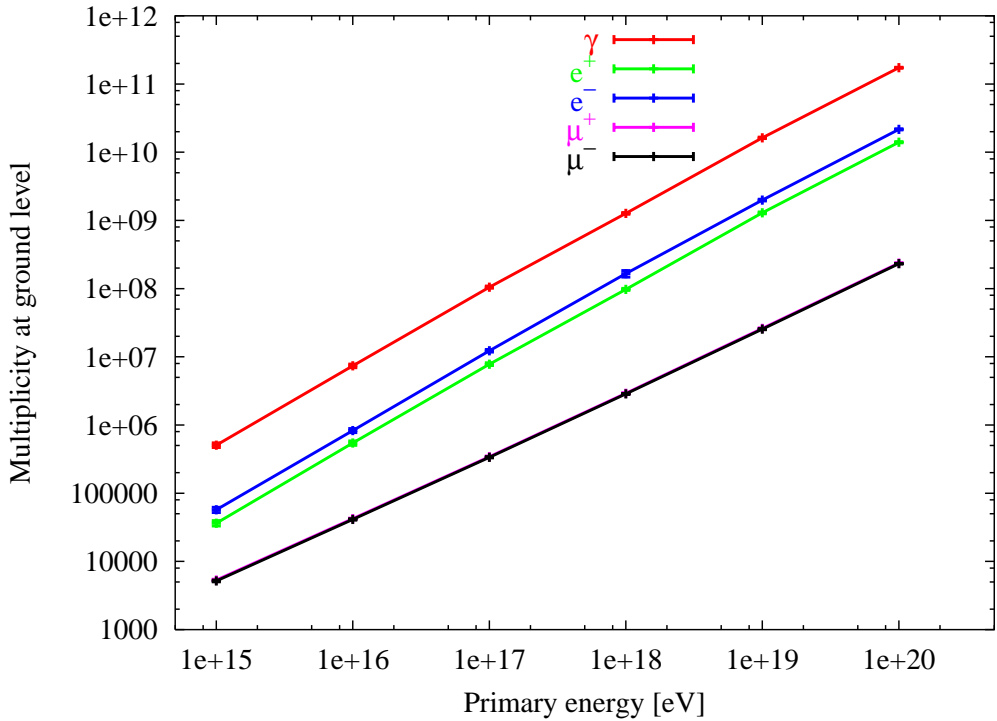}}} \par}
\caption{Multiplicities of photons, \protect\( e^{\pm }\protect \), \protect\( \mu ^{\pm }\protect \)
at the ground level as a function of the primary energy. Due to the
logarithmic scale, \protect\( \mu ^{+}\protect \) and \protect\( \mu ^{-}\protect \)
look superimposed.}
\label{third}
\end{figure}

\section{Structural Properties of the shower}
In ref. \cite{ccf} we have described some of the features 
of the cascade induced by cosmic rays (protons), 
performing some large scale simulation of the 
shower formation both with a fixed and a variable first impact 
- at zero zenith - of the primary.  
The arrival direction is therefore perpendicular to the detector plane and 
the opening is measured with respect to this vertical axis.
 Here we intend to describe some of the observables of the shower which may serve as a characterization of its structure. 

We start by showing some results obtained using the Monte Carlo 
air shower generator CORSIKA. The simulations have been performed on a small cluster 
running $openmoses$ for parallelization purposes and 
we have used the thinning option 
in the compiled code in order to make our results manageable. Even with these 
simplifications, the number of possible channels at the GZK energy generate 
total outputs around 1 Terabyte of memory in size. 

Fig.~1 illustrates a simulation of the 
multiplicites as a function of the various observation 
levels in the atmosphere. The impact of the primary 
has been assumed to take place almost 
at the beginning (113 Km) of the atmosphere 
and the developement of the multiplicities of the 
main components (photons, electrons, muons and antiparticles) 
has been tracked. Unfortunately the statistical 
fluctuations in this results (not shown in the figure, see \cite{cacf}) 
are rather large and considering the rate at which AUGER will presumably collect UHECR 
events (30 to 100 per year), we will probably be able 
to confront the experimental 
data with simulation of this type only after few years of run ($\sim 4$).

\section{Inclusive Multiplicities}
While fluctuations in the geometrical opening of the shower 
can be rather wide, those in the multiplicities 
are rather reduced \cite{cacf}.
Therefore, the study of the inclusive multiplicities of the various subcomponents of the air shower as a function of the energy or as a function 
of the height in the atmosphere are 
particularly interesting and deserve some comments. 
Distributions such as those in Fig.~2 are 
a good characterization of the multiplicities of the subcomponents of 
the shower across the various observation levels in the atmosphere and contain important 
information useful for its reconstruction. 

Fig.~3, instead, shows that simulation models such as CORSIKA, combined with 
hadronization models - in the first hadronic interaction - such 
as QGSJET \cite{QGSJET} predict linearly rising multiplicities (with energy, in a log/log plot) 
for the various components. In regard to this, 
it a has been shown in \cite{cacf} that the total multiplicities 
do not seem to be affected appreciably 
by the statistical fluctuations in the formation of the shower, 
and therefore 
they can be taken as a robust characterization both of the hadronization 
model and of the overall developement of the cascade. 

Given the robustness of this result, a 
failure to reproduce experimentally this linear trend can be considered 
either as a serious fault of the models implemented so far or 
as a possible signal of new physics. 

\section{Calibrations}
An interesting possibility to take into consideration is the production of dark matter 
as a consequence of the first (primary) impact. This may be the case if 
supersymmetric channels open up at the large energy scale of the first collision of 
the primary, with a subsequent production of neutralinos or other weakly 
interacting dark matter candidates. If the underlying interaction 
favors channels with large missing energy, then the linear trend 
shown in Fig.~2 
can be affected. 
Given the fact that new interactions are likely to open up at a higher energy,
the experimental calibration (the slopes and intercepts) 
could be done at a lower energy, say below the cutoff, and deviations from 
this
linear behaviour measured above it. Notice that these measurements are properties of a 
single shower and as such are not affected by the variations of the flux or of all the other 
parameters characterizing the inclusive high energy cosmic rays spectrum. Only the 
particle type of the primary is supposed to be held fixed 
(a proton, in this case).

 One of the 
issues that need to be addressed is how to characterize the shower in such a way that a reduced measured multiplicity of an event, due to missing energy, 
is not interpreted as an event of lower energy. 
To avoid this, we think that one needs a combined study of the geometry 
of the modified shower and of the lateral distributions. 
If the missing energy takes place mostly at the beginning of the shower, this might be possible, since the geometry 
of the upper part of the shower could be affected. 
The use of fluorescence detectors may be able to reconstruct this anomaly.
Instead, for a uniform multiplicity loss along the entire atmosphere 
the shower of reduced multiplicity is unlikely to be correctly identified 
as such and will probably be confused for an event of lower energy. 
However, as the original energy, available at the first impact, 
degrades due to scattering and fragmentation, there should 
be a height at which the shower shows an anomaly, 
since supersymmetric channels are not available any longer. 

\section{Conclusions}
Searches for the origin of the UHECR will span probably a decade, 
but AUGER is already collecting data as the detector tanks are deployed on the site. It will be crucial, in order to confront theory with experiment, 
to have improved models of the hadronic interaction, 
which may also include possible extensions of the standard model in the Monte Carlo reconstruction 
of the topology and multiplicity of the event. We have pointed out that the issue 
of missing multiplicities could be relevant to identify new physics. 
Large missing energy events (probably around 20 \%) 
may have a chance to be identified, with important implications.

\centerline{\bf Acknowledgements}
We thank Alon Faraggi for collaborating to this analysis
and to J. Knapp for very informative discussions about AUGER 
while completing this work. We also thank INFN for supporting 
the building of a small cluster at the Univ. of Lecce dedicated to these studies.

\end{document}